\documentclass[preprint,aps,nofootinbib,showpacs,amsfonts,epsf]{revtex4}

\input{epsf.tex}

\newcommand{\E}{{\cal{E}}}
\newcommand{\s}{\sigma}

\renewcommand{\a}{\alpha}
\renewcommand{\k}{\kappa}

\newcommand{\be}{\begin{equation}}
\newcommand{\ee}{\end{equation}}
\newcommand{\bea}{\begin{eqnarray}}
\newcommand{\eea}{\end{eqnarray}}
\newcommand{\ba}{\begin{array}}
\newcommand{\ea}{\end{array}}
\def\J#1#2#3#4{{#1} {\bf #2}, #3 (#4)}
\def\PRD{Phys. Rev. D}
\def\PR{Phys. Rev.}
\def\PRL{Phys. Rev. Lett.}
\def\PTP{Prog. Theor. Phys.}

\def\APL{Ann. Phys. (Leipzig)}

\def\JMP{J. Math. Phys.}

\def\CMP{Commun. Math. Phys.}

\def\CQG{Class. Quantum Grav.}

\def\PLA{Phys. Lett. A}
\def\PLB{Phys. Lett. B}

\def\JHEP{J. High Energy Phys.}

\begin{document}
\draft
\title{Equatorially symmetric configurations\\ of two Kerr-Newman black holes}

\author{V.~S.~Manko$^\dagger$ and E.~Ruiz$^\ddagger$}
\address{$^\dagger$Departamento de F\'\i sica, Centro de Investigaci\'on y
de Estudios Avanzados del IPN, A.P. 14-740, 07000 Ciudad de
M\'exico, Mexico\\$^\ddagger$Instituto Universitario de F\'{i}sica
Fundamental y Matem\'aticas, Universidad de Salamanca, 37008
Salamanca, Spain}

\begin{abstract}
In this paper, we employ the general equatorially symmetric
two-soliton solution of the Einstein-Maxwell equations for
elaborating two physically meaningful configurations describing a
pair of equal Kerr-Newman corotating black holes separated by a
massless strut. The first configuration is characterized by
opposite magnetic charges of its constituents, while in the second
configuration the black holes carry equal electric and opposite
magnetic charges, thus providing a nontrivial example of a binary
dyonic black-hole system. The thermodynamic properties of these
binary configurations are studied and the first law of
thermodynamics taking correctly into account the magnetic field
contribution is formulated for each case.
\end{abstract}

\pacs{04.20.Jb, 04.70.Bw, 97.60.Lf}

\maketitle

\section{Introduction}

In the paper \cite{MMR}, the general six-parameter two-soliton
solution of the Einstein-Maxwell equations possessing equatorial
symmetry was constructed with the aid of Sibgatullin's integral
method \cite{Sib}. It is able to describe the exterior
gravitational and electromagnetic fields of compact objects, as
well as of the binary systems of identical black holes or
hyperextreme sources. While the former application of that
solution is better known in the literature (see, e.g.,
Refs.~\cite{MRu1,MRu2}), the latter possibility of solution's
usage for the analysis of the black hole binary systems has been
scantly exploited only recently in its pure vacuum sector, and
therefore it would be certainly of interest to make use of the
solution \cite{MMR} (henceforth referred to as the MMR solution)
for obtaining its physically interesting generic electrovacuum
subfamilies representing two equal (up to the sign of the charges)
Kerr-Newman (KN) black holes \cite{New} separated by a massless
strut \cite{Isr}. The main objective of the present paper will be
derivation and analysis of a nontrivial binary configuration of
dyonic KN black holes carrying equal electric and opposite
magnetic charges and formulation for it of the first law of
thermodynamics.

Though our main results which will be discussed in the present
paper were obtained more than a year ago, their publication was
postponed due to the paper \cite{CGa} criticizing the extension of
the well-known Smarr mass formula \cite{Sma} to the case of dyonic
black holes \cite{MGa}, and the criticism has been refuted only
recently \cite{GMR} by demonstrating that the model of the dyonic
KN solution worked out in \cite{CGa} was frankly unphysical.
However, lately an effort has been made \cite{Cab} to use the
binary configurations of dyonic KN black holes for rehabilitating
the approach of the paper \cite{CGa} to the Smarr formula, so we
find it necessary and instructive to briefly comment in the
discussion section of our paper on the contradictions of the
preprint \cite{Cab}.

The plan of the present paper is as follows. In the next section
we will write down the MMR solution in a form simpler than the
original one thanks to some technical improvements in the
construction procedure that have been found over the years. This
representation is fundamental for the subsequent working out the
particular and generic cases of our interest. In Sec.~III we
consider a binary configuration of corotating KN black holes
endowed with opposite magnetic charges. This particular binary
system will permit us to present the corresponding magnetic
version of the Smarr formula and show that it practically does not
differ from the usual mass relation in the case of opposite
electric charges. Here we also derive the first law of
thermodynamics for that binary system and find the corresponding
expression of the thermodynamic length. In Sec.~IV we show how the
binary configuration of corotating KN black holes with equal
electric and opposite magnetic charges is contained in the general
MMR solution and analyze its thermodynamical properties, including
the corresponding first law of thermodynamics and correct account
for the magnetic contribution in it. Discussion of the results
obtained and concluding remarks can be found in Sec.~V, where in
particular we touch an interesting question of the nonuniqueness
of the binary systems of KN sources with the same masses, angular
momenta and charges.

\section{Enhanced form of the MMR solution}

The MMR solution was constructed from the expressions of the Ernst
complex potentials \cite{Ern} on the upper part of the symmetry
axis (the axis data) of the form
\be e(z)=\frac{(z-m-ia)(z+ib)+k} {(z+m-ia)(z+ib)+k}, \quad
f(z)=\frac{qz+ic} {(z+m-ia)(z+ib)+k}, \label{AD1} \ee
where six arbitrary real parameters $m$, $a$, $b$, $k$, $q$ and
$c$ are related to the first two nonzero mass, angular momentum,
electric and magnetic multipoles \cite{Sim,HPe,SAp} by the
formulas
\bea M_0&=&m, \quad M_2=-m(k+a^2), \quad J_1=ma, \quad
J_3=-m[k(2a-b)+a^3], \nonumber\\ Q_0&=&q, \quad
Q_2=-q(k+b^2)-(a-b)(c+aq), \quad B_1=c+q(a-b), \nonumber\\
B_3&=&-c(k+b^2)-(a-b)[q(a^2+b^2+2k)+ac]. \label{mm} \eea

The position of the sources on the symmetry axis is defined by
four roots $\a_i$ of the algebraic equation
\be e(z)+\bar e(z)+2f(z)\bar f(z)=0 \label{SE} \ee
(a bar over a symbol means complex conjugation), and so $\a_i$
have the form
\bea \a_1&=&-\a_4=\frac{1}{2}(\kappa_++\kappa_-), \quad
\a_2=-\a_3=\frac{1}{2}(\kappa_+-\kappa_-), \nonumber\\
\kappa_\pm&=&\sqrt{m^2-a^2-b^2-q^2-2k\pm2d}, \quad
d=\sqrt{(k+ab)^2-m^2b^2+c^2}. \label{alfas1} \eea

The form of the Ernst potentials $\E$ and $\Phi$ in the whole
($\rho,z$) space obtainable from the axis data (\ref{AD1}) is
given by the expressions
\bea \E&=&(A-B)/(A+B), \quad \Phi=C/(A+B), \nonumber\\
A&=&\k_+^2\{[(d-ab-k)\k_-^2+k(m^2-q^2)-(aq+c)(bq-c)]
(R_+r_-+R_-r_+) \nonumber\\
&&+i\k_-[(a-b)(ab+k-d)-m^2b+qc](R_+r_--R_-r_+)\} \nonumber\\
&&+\k_-^2\{[(d+ab+k)\k_+^2-k(m^2-q^2)+(aq+c)(bq-c)]
(R_+r_++R_-r_-) \nonumber\\
&&-i\k_+[(a-b)(ab+k+d)-m^2b+qc](R_+r_+-R_-r_-)\} \nonumber\\
&&-4d[[k(m^2-q^2)-(aq+c)(bq-c)](R_+R_-+r_+r_-), \nonumber\\
B&=&m\k_+\k_-\{d[\k_+\k_-(R_++R_-+r_++r_-) -(m^2-a^2+b^2-q^2)
(R_++R_--r_+-r_-)] \nonumber\\ &&+ibd[(\k_++\k_-)(R_+-R_-)
+(\k_+-\k_-)(r_--r_+)] \nonumber\\ &&+i[b(m^2-a^2)-ak-qc]
[(\k_++\k_-)(r_+-r_-)+(\k_+-\k_-)(R_--R_+)]\}, \nonumber\\
C&=&qB/m +\k_+\k_-(bq-c)[2d(b-a)(R_++R_--r_+-r_-) \nonumber\\
&&-i\k_+(d+ab+k)(R_+-R_--r_++r_-)
-i\k_-(d-ab-k)(R_+-R_-+r_+-r_-)], \nonumber\\
R_\pm&=&\sqrt{\rho^2+\left[ z\pm\frac{1}{2}
(\kappa_++\kappa_-)\right]^2}, \quad r_\pm=\sqrt{\rho^2+\left[
z\pm\frac{1}{2} (\kappa_+-\kappa_-)\right]^2}, \label{EF1} \eea
and these formulas are presented in a simpler form than in the
original paper \cite{MMR}. The corresponding metric functions $f$,
$\gamma$ and $\omega$ from the Weyl-Papapetrou stationary
axisymmetric line element \cite{Pap}
\be d s^2=f^{-1}[e^{2\gamma}(d\rho^2+d z^2)+\rho^2 d\varphi^2]-f(d
t-\omega d\varphi)^2 \label{Papa} \ee
have the following form:
\bea f&=&\frac{A\bar A-B\bar B+C\bar C}{(A+B)(\bar A+\bar B)},
\quad e^{2\gamma}=\frac{A\bar A-B\bar B+C\bar C}{16d^2\k_+^4
\k_-^4R_+R_-r_+r_-}, \quad \omega=-\frac{{\rm Im}
[G(\bar A+\bar B)+C\bar I]}{A\bar A-B\bar B+C\bar C}, \nonumber\\
G&=&-2(z-ia)B-qC+\k_+\k_-^2[d(2m^2-q^2)-2m^2b^2+c^2]
(R_-r_--R_+r_+) \nonumber\\
&&+\k_+^2\k_-\{[d(2m^2-q^2)+2m^2b^2-c^2] (R_-r_+-R_+r_-)
-i\k_-(2m^2b-qc)(R_+-R_-) \nonumber\\ &&\times (r_+-r_-)\}
+i\{(a-b)[k(2m^2-q^2)-(aq+c)(bq-c)]-m^2q(bq-c)\} \nonumber\\
&&\times[\k_-^2(R_+r_++R_-r_-)-\k_+^2(R_+r_-+R_-r_+)
+4d(R_+R_-+r_+r_-)]+m\k_+\k_- \nonumber\\ &&\times
\{2k[\k_-(d-ab-k) (R_-+r_--R_+-r_+)+\k_+(d+ab+k)
(R_--r_--R_++r_+)] \nonumber\\ &&+c(c-bq)[(\k_+-\k_-)(R_--R_+)
-(\k_++\k_-)(r_--r_+)] \nonumber\\ &&+2id[2k(a-b)+q(c-bq)]
(R_--r_-+R_+-r_+)\}, \nonumber\\ I&=&q(A+B)+icB/m- [z-i(a-b)]C
+m\k_+\k_-[\k_-(dq-bc)(R_-r_--R_+r_+) \nonumber\\
&&+\k_+(dq+bc)(R_-r_+-R_+r_-) -2id(bq-c)(R_+R_--r_+r_-
+2\k_+\k_-)] \nonumber\\ &&-imd(bq+c)[\k_-^2(R_+r_++R_-r_-)
+\k_+^2(R_+r_-+R_-r_+)]+im[bq(m^2-a^2) \nonumber\\
&&-c(b^2+q^2)-k(aq+c)] [\k_-^2(R_+r_++R_-r_-)
-\k_+^2(R_+r_-+R_-r_+)] \nonumber\\ &&-2imd[(bq-c)
(m^2-a^2+b^2+q^2)-2kq(a-b)](R_+R_-+r_+r_-)-\k_+\k_- \nonumber\\
&&\times\{\k_+[(d+ab+k)(ac-abq-kq)+2m^2b(bq-c)] (R_--r_--R_++r_+)
\nonumber\\
&&+\k_-[(d-ab-k)(ac-abq-kq)-2m^2b(bq-c)](R_-+r_--R_+-r_+)
\nonumber\\ &&+2id[(a-b)(ac-abq-kq)+2m^2(bq-c)]
(R_--r_-+R_+-r_+)\},
 \label{MF1} \eea
while the nonzero components of the electromagnetic four-potential
are defined as
\be A_t=-{\rm Re}\left(\frac{C}{A+B}\right), \quad A_\varphi={\rm
Im}\left(\frac{I}{A+B}\right). \label{At} \ee

It may be noted that the expression of the metric function
$\omega$ is determined by only two additional potentials $G$ and
$I$, in contradistinction to the three such potentials in the
original paper \cite{MMR}, which obviously improves the
presentation of the MMR solution.

Due to its multipole structure (\ref{mm}) involving physically
important multipole moments, the MMR metric is able to describe
the exterior field of compact massive objects endowed with
electric charge and magnetic dipole moment, and in this relation
its most recent application was considered in the paper
\cite{MRu2}. At the same time, the above formulas also contain, as
special subfamilies, the solutions for two equal corotating KN
sources, black holes or hyperextreme objects, and we now turn to
the consideration of these binary configurations, mainly
concentrating on the black-hole systems.

\section{Two equal corotating KN black holes with opposite charges}

After the publication of our work on two corotating identical Kerr
sources \cite{MRu3} it was of course logic for us to turn our
attention to searching the analogous binary equatorially symmetric
configurations of KN sub- and hyperextreme constituents. It
appears that the MMR solution provides the simplest way to
identify and describe the latter configurations because these
arise from the formulas of the previous section by just imposing
the condition $\omega=0$ on the intermediate part of the symmetry
axis (the axis condition). While treating the problem of two KN
sources separated by a massless strut, it is advantageous to
reparametrize the quantities $\a_i$ and the axis data (\ref{AD1})
in the form
\be \a_1=-\a_4=\frac{1}{2}R+\sigma, \quad
\a_2=-\a_3=\frac{1}{2}R-\sigma, \label{alfas} \ee
and
\bea e(z)=\frac{z^2-2(m+ia)z-{\textstyle\frac14}R^2
+2(m^2-a^2-q^2)-\s^2+i\delta} {z^2+2(m-ia)z-{\textstyle\frac14}R^2
+2(m^2-a^2-q^2)-\s^2-i\delta}, \nonumber\\
f(z)=\frac{2qz+ib} {z^2+2(m-ia)z-{\textstyle\frac14}R^2
+2(m^2-a^2-q^2)-\s^2-i\delta}, \label{axis1} \eea
with
\be
\delta=\epsilon\sqrt{(\s^2-m^2+a^2+q^2)[R^2-4(m^2-a^2-q^2)]+b^2},
\quad \epsilon=\pm1, \label{delta1} \ee
where the set of six arbitrary parameters is now comprised of $m$,
$a$, $R$, $\s$, $q$ and $b$. Note that the idea of the
reparametrization consists in introducing the roots of equation
(\ref{SE}) explicitly into the axis data, and one can see that
formulas (\ref{axis1})-(\ref{delta1}) reduce to the axis data for
two equal corotating Kerr sources considered in \cite{MRu3} in the
vacuum limit ($q=b=0$). The substitution that casts the axis data
(\ref{AD1}) into the form (\ref{axis1}) is the following:
\bea &&m\to2m, \quad a\to\frac{4ma-\delta}{2m}, \quad
b\to-\frac{\delta}{2m}, \quad
k\to\frac{4m(ms+a\delta)-\delta^2}{4m^2}, \quad q\to2q, \quad c\to
b, \nonumber\\ &&s\equiv 2(m^2-a^2-q^2)-\frac{1}{4}R^2-\s^2.
\label{sub1} \eea
We also notice that for some calculations it may be advantageous
to use $\delta$ as an arbitrary parameter, in which case the
expression of $\s$ in terms of $\delta$ following from
(\ref{delta1}) has the form
\be \s=\sqrt{m^2-a^2-q^2+\frac{\delta^2-b^2}{R^2-4(m^2-a^2-q^2)}}.
\label{sigma} \ee

The subfamily of the MMR spacetime representing two equal KN
sources separated by a massless strut is segregated from the
general case by the condition
\be \omega\left(\rho=0,|z|\le\textstyle{\frac12}R-{\rm Re}
(\s)\right)=0, \label{AC} \ee
which ensures that the constituents do not overlap. The quickest
way to get the explicit form of (\ref{AC}) is to use the formulas
for $\omega$ from the previous section and in the axis expression
of $\omega$ calculated for $\rho=0$, $|z|\le\a_2$ to perform the
parameter change (\ref{sub1}) supplemented with the substitutions
\be \kappa_+\to R, \quad \kappa_-\to2\s, \quad d\to(R^2-4\s^2)/4.
\label{sub2} \ee
Unlike in the vacuum case of corotating Kerr sorces \cite{MRu3}
where the condition (\ref{AC}) results in a quadratic equation for
the quantity $\s$, in the case of the reparametrized MMR solution
the axis condition leads to a biquadratic equation for $\s$ that
can be readily solved yielding
\bea \s^2&=&\frac{1}{32a^2(R^2+2mR+4a^2+2q^2)^2}\Bigl(
-D+2(R^2+2mR+4a^2+2q^2) \nonumber\\ &&\times\{(R^2-4m^2+4a^2+4q^2)
[(R^2+2mR+4q^2)^2+4a^2(R^2+4m^2-4a^2+4q^2) \nonumber\\ &&+24aqb]
+8a [(mR+2m^2-q^2)(R^2a+4qb)+ab^2]\} \nonumber\\
&&\pm [(R+2m)(R^2+2mR+4a^2+4q^2)+8ma^2]\sqrt{D(R^2+4mR+4m^2+4a^2)}
\,\Bigr), \nonumber\\ D&=&(R^2-4m^2+4a^2+4q^2)^2 [(R^2+2mR+4q^2)^2
+4a^2(R^2+8q^2)] \nonumber\\ &&+32ab(R^2+2mR+4a^2+2q^2)
[q(R^2-4m^2+4a^2+4q^2)+ab], \label{sig1} \eea
and this determines the subfamily of equatorially symmetric
configurations of two KN sources, black holes or naked
singularities, kept apart from each other by a massless strut.

With the reparametrization made, the moments $M_0$, $J_1$, $Q_0$
and $B_1$ defining, respectively, the total mass, total angular
momentum, total charge and magnetic dipole moment of the binary
system take the form
\be M_0\equiv M_T=2m, \quad J_1\equiv J_T=4ma-\delta, \quad
Q_0\equiv Q_T=2q, \quad B_1\equiv\mu=4aq+b, \label{mm2} \ee
so that we can further precise the interpretation of the KN
sources in the subfamily (\ref{sig1}) as carrying equal electric
and opposite magnetic charges.

The particular case for which one might expect simplification of
the expression of $\s$ in (\ref{sig1}) is the absence of electric
charges ($q=0$), when only two opposite magnetic charges are
present. In what follows we shall elaborate this case in more
detail, restricting our analysis exclusively to the black-hole
configurations corresponding to real valued $\s$. The case of
nonzero $q$ and vanishing magnetic charges constitutes a
specialization of the general subfamily of binary systems which
will be considered later, and one may recall in this respect that
the electric charge in principle can be readily introduced into a
binary system of Kerr black holes via the well-known
Ernst-Harrison transformation \cite{Ern,Har}.

After setting $q=0$, the axis data (\ref{AD1}) take the form
\bea e(z)&=&\frac{z^2-2(m+ia)z-{\textstyle\frac14}R^2
+2(m^2-a^2)-\s^2+i\delta} {z^2+2(m-ia)z-{\textstyle\frac14}R^2
+2(m^2-a^2)-\s^2-i\delta}, \nonumber\\
f(z)&=&\frac{ib} {z^2+2(m-ia)z-{\textstyle\frac14}R^2
+2(m^2-a^2)-\s^2-i\delta}, \label{axis2} \eea
and for our purposes we must write down the corresponding Ernst
potentials and metric functions using the formulas of the previous
section together with the substitutions (\ref{sub1}) and
(\ref{sub2}). In this way we obtain for the Ernst potentials the
expressions
\bea \E&=&(A-B)/(A+B), \quad \Phi=C/(A+B), \nonumber\\
A&=&R^2\{[4m^2(m^2-2\s^2)-(R^2+4a^2)(a^2-\s^2)+4ma\delta]
(R_+r_-+R_-r_+) \nonumber\\
&&-2i\s[a(R^2-4m^2+4a^2)-2m\delta](R_+r_--R_-r_+)\} \nonumber\\
&&+4\s^2\{[2m^2(R^2-2m^2)-(R^2-4a^2)(a^2+\s^2)-4ma\delta]
(R_+r_++R_-r_-) \nonumber\\
&&-2iR[2a(m^2-a^2-\s^2)+m\delta](R_+r_+-R_-r_-)\} \nonumber\\
&&+(R^2-4\s^2)[(R^2+4a^2)(a^2+\s^2)-4m^4-4ma\delta](R_+R_-+r_+r_-), \nonumber\\
B&=&2R\s\{(R^2-4\s^2)[mR\s(R_++R_-+r_++r_-) -(2m^3-2ma^2+a\delta)
(R_++R_- \nonumber\\ &&-r_+-r_-)]
+i[ma(R^2+4\s^2-8m^2+8a^2)-2(m^2+a^2)\delta][(R-2\s)(R_--R_+) \nonumber\\
&&+(R+2\s)(r_+-r_-)]  +iR\s\delta
[(R-2\s)(R_--R_+)-(R+2\s)(r_+-r_-)]\}, \nonumber\\
C&=&2Rb\s[a(R^2-4\s^2)(R_++R_--r_+-r_-)
+2iR(m^2-a^2-\s^2)(R_+-R_--r_++r_-) \nonumber\\
&&+i\s(R^2-4m^2+4a^2)(R_+-R_-+r_+-r_-)], \nonumber\\
R_\pm&=&\sqrt{\rho^2+\left[z\pm({\textstyle{\frac{1}{2}}}R+\s)\right]^2},
\quad
r_\pm=\sqrt{\rho^2+\left[z\pm({\textstyle{\frac{1}{2}}}R-\s)\right]^2},
\label{EF2} \eea
while for the metric functions we get
\bea f&=&\frac{A\bar A-B\bar B+C\bar C}{(A+B)(\bar A+\bar B)},
\,\,\, e^{2\gamma}=\frac{A\bar A-B\bar B+C\bar C}{16R^4\s^4
(R^2-4\s^2)^2R_+R_-r_+r_-}, \,\,\, \omega=-\frac{{\rm Im}
[G(\bar A+\bar B)+C\bar I]}{A\bar A-B\bar B+C\bar C}, \nonumber\\
G&=&4R\s^2[2(R^2+4a^2)(2m^2-a^2-\s^2) -8m^4-b^2]
(R_-r_--R_+r_+)\nonumber\\
&&+2R^2\s\{
[2(R^2-8m^2+4a^2)(a^2+\s^2)+8m^4+b^2] (R_-r_+-R_+r_-) \nonumber\\
&&+8im\delta\s(R_+-R_-)(r_+-r_-)\}
+2ia[2(R^2+4a^2)(a^2+\s^2)-8m^4+b^2-8ma\delta] \nonumber\\
&&\times[R^2(R_+-r_+)(r_--R_-)-4\s^2(R_+-r_-)(r_+-R_-)]-2R(\s/m)
\nonumber\\ &&\times \{[(R^2+4a^2)(a^2+\s^2)-4m^4+b^2
-4ma\delta][2R(m^2-a^2-\s^2) (R_--R_+-r_-+r_+) \nonumber\\
&&+\s(R^2-4m^2+4a^2) (R_--R_++r_--r_+)
+ia(R^2-4\s^2)(R_-+R_+-r_--r_+)] \nonumber\\
&&-2m^2b^2[(R-2\s)(R_--R_+)-(R+2\s)(r_--r_+)]\}
-2zB+i(4a-\delta/m)B, \nonumber\\
I&=&2Rb\s[R\delta(R_+r_--R_-r_+)
-2\delta\s(R_+r_+-R_-r_-) +im(R^2-4\s^2)(R_+R_- \nonumber\\
&&-r_+r_-+4R\s)] -(i/2)mB_0(R^2-4\s^2)[R^2(R_+r_-+R_-r_+)
+4\s^2(R_+r_++R_-r_-)]
\nonumber\\
&& +(i/2)b[m(R^2-8m^2+8a^2+4\s^2)-4a\delta] [R^2(R_+r_-+R_-r_+)
\nonumber\\
&&-4\s^2(R_+r_++R_-r_-)] +2ib(R^2-4\s^2)[2m(m^2-a^2)+a\delta]
(R_+R_-+r_+r_-) \nonumber\\
&&-Rb(\s/m)\{2R[(4ma-\delta)(m^2-a^2-\s^2)+4m^2\delta]
(R_--R_+-r_-+r_+)
\nonumber\\
&&+\s[(4ma-\delta)(R^2-4m^2+4a^2)-16m^2\delta](R_--R_++r_--r_+)
-i(R^2-4\s^2) \nonumber\\
&&\times(8m^3-4ma^2+a\delta)(R_-+R_+-r_--r_+)\}
+ibB/(2m)-(z-2ia)C. \label{MF2} \eea
Formulas (\ref{At}) for the electromagnetic potentials $A_t$ and
$A_\varphi$ do not change.

The key point in the simplification of $\s$ in (\ref{sig1}) is
finding the form of the parameter $b$ in terms of the individual
magnetic charge $\beta$ of one of the black-hole constituents
determined by the formula
\be \beta=\frac{1}{2}\int_H \omega A_{t,z}dz, \label{mch} \ee
where both functions $\omega$ and $A_t$ in (\ref{mch}) must be
evaluated on the horizon. Of course, in view of the equatorial
symmetry of our binary configuration it is sufficient to calculate
the physical characteristics of only one of the black holes. The
tedious but straightforward calculations eventually lead to the
following rather simple relation
\be b=-\frac{\beta R(R^2-4m^2+4a^2)[(R+2m)^2+4a^2]}
{(R^2+2mR+4a^2)[(R+2m)^2+4a^2]-8a^2\beta^2}, \label{b1} \ee
where $\beta$ is the magnetic charge of the lower black hole whose
horizon is the rod $-\textstyle{\frac{1}{2}}R-\s\le z\le
-\textstyle{\frac{1}{2}}R+\s$ located on the symmetry axis (see
Fig.~1), the magnetic charge of the upper constituent being
$-\beta$.

The substitution of (\ref{b1}) into (\ref{sig1}) converts the
radicand in the latter formula into a perfect square, so that
choosing in (\ref{sig1}) the minus sign we arrive at the final
expression for $\s$ in the form
\be \s=\sqrt{m^2-a^2+\frac{(R^2-4m^2+4a^2)
\{4a^2[m-\beta_0(R+2m)]^2-R^2\beta^2\}}
{[R^2+2mR+4a^2(1-2\beta_0)]^2}}, \label{sig2} \ee
while for $\delta$, taking into account (\ref{b1}) and
(\ref{sig2}), we get from (\ref{delta1})
\be \delta=\frac{2a(R^2-4m^2+4a^2)[m-\beta_0(R+2m)]}
{R^2+2mR+4a^2(1-2\beta_0)}, \quad \beta_0\equiv
\frac{\beta^2}{(R+2m)^2+4a^2}, \label{delta2} \ee
where we have introduced a dimensionless parameter $\beta_0$ for
writing down the results in a more concise form.

Since the magnetically charged KN black holes are equal and
corotating, their individual Komar \cite{Kom} masses and angular
momenta are just halves the respective total quantities $M_T$ and
$J_T$, so that $m$ is the mass of each black hole, and for the
individual angular momenta $J$ we get from (\ref{mm2}) and
(\ref{delta2})
\be J=\frac{a[(R+2m)^2+4a^2][m+\beta_0(R-2m)]}
{R^2+2mR+4a^2(1-2\beta_0)}. \label{AM1} \ee
The other physical characteristics that might be of interest to us
are the horizon's area ${\cal A}$, the surface gravity $\kappa$,
horizon's angular velocity $\Omega$ and the magnetic potential
$\Phi_m$, which all can be calculated by means of the formulas of
the paper \cite{Tom}, taking into account the relation of $\Phi_m$
to the electric potential $\Phi_e$ of the associated problem
\cite{MGa}. Assuming the validity of the Bekenstein-Hawking
formula $S={\cal A}/4$ between the entropy $S$ and horizon's area
${\cal A}$ \cite{Bek,Haw}, and also recalling that the Hawking
temperature $T$ is related to the surface gravity as
$T=\kappa/(2\pi)$, we give below the formulas for $S$, $T$,
$\Phi_m$ and $\Omega$ calculated for the lower black hole of our
particular binary configuration:
\bea S&=&\frac{\s}{2T}= \frac{\pi[(R+2m)^2+4a^2]\lambda_0}
{(R+2\s)[R^2+2mR+4a^2(1-2\beta_0)]}, \nonumber\\
\Omega&=&\frac{a\nu_0} {[(R+2m)^2+4a^2]\lambda_0}, \nonumber\\
\Phi_m&=&\frac{\beta(R^2-4m^2+4a^2)[(R+2m)(m+\s)-2a^2]}
{[(R+2m)^2+4a^2]\lambda_0}, \label{TQ1} \eea
where
\bea \lambda_0&=&2m[(R+2m)(m+\s)-2a^2]-\beta_0
[(R+2m)(R^2-4m^2)+8a^2(R+m+\s)], \nonumber\\
\nu_0&=&[R^2+2\s(R+2\s)-4m^2+4a^2][R^2+2mR+4a^2(1-2\beta_0)]
\nonumber\\ &&-4m(R^2-4m^2+4a^2)[m-(R+2m)\beta_0], \label{ln0}
\eea
and these thermodynamical quantities verify the Smarr mass formula
\cite{Sma}
\be m=2TS+2\Omega J+\Phi_m\beta, \label{Sma} \ee
which also holds for the upper black hole whose magnetic potential
is $-\Phi_m$ and magnetic charge $-\beta$.

To the above thermodynamic characteristics we must add the
expressions of the interaction force ${\cal F}$ \cite{Isr} and
thermodynamic length $\ell$ \cite{AGK} which were shown to enter
explicitly into the first law of thermodynamics in the static and
stationary vacuum \cite{HKK,HRR,AGK,RGM} and electrovacuum cases
\cite{KZe,GMR2}. It is remarkable that both ${\cal F}$ and $\ell$
are defined in terms of the value $\gamma_0$ of the metric
function $\gamma$ on the strut, and whereas the formula for ${\cal
F}$ is well known, the analogous formula for $\ell$, namely,
$\ell=L\exp(\gamma_0)$, where $L$ is the coordinate length of the
strut, has been discovered only recently \cite{KZe}. The form of
${\cal F}$ and $\ell$ in our case has been found to be
\bea {\cal F}&=&\frac{[(R+2m)^2-4a^2](m^2-4a^2\beta_0^2)
+\beta_0[R^2(R+2m)^2+16a^2(m^2-a^2)]}
{(R^2-4m^2+4a^2)[(R+2m)^2+4a^2]}, \nonumber\\
\ell&=&\frac{(R^2-4m^2+4a^2)^2[(R+2m)^2+4a^2]}
{(R+2\s)[R^2+2mR+4a^2(1-2\beta_0)]^2}, \label{Fl1} \eea
so that the corresponding first law of thermodynamics can be
written, following the procedure described in \cite{KZe,RGM}, in
the form
\be dM_T=2TdS+2\Omega dJ+2\Phi_m d\beta-\ell d{\cal F}, \quad
M_T=2m. \label{FL1} \ee

It is worth noting that the case of corotating KN black holes with
opposite electric charges is trivially obtainable from the above
configuration of magnetically charged KN black holes by formally
changing in (\ref{EF2}) the electromagnetic Ernst potential $\Phi$
to $i\Phi$, in which case $b$ becomes an electric dipole
parameter, while $\beta$ becomes the electric charge. Moreover,
the transformation $b\to b-ip$, $b^2\to b^2+p^2$ in the formulas
(\ref{EF2}), (\ref{FL1}) and (\ref{b1}) leads, after the analogous
complex extension of the magnetic charge parameter
$\beta\to\beta-iq$, $\beta^2\to \beta^2+q^2$ to the case of two
dyonic KN black holes endowed with opposite electric and magnetic
charges, and then the Smarr mass relation takes the form discussed
in \cite{MGa}. In the paper \cite{MRS} it was clarified that in
order to treat correctly the solutions involving both electric and
magnetic charges it is best to identify first the particular case
in which only the electric charges are present and then apply the
extension parameter procedure. Our purely magnetic solution
considered in this section illustrates well that the solution with
solely magnetic charges is equally suitable as a starting point
for consistently treating the more general cases.

\section{Two corotating dyonic KN black holes with equal
electric and opposite magnetic charges}

We now turn to the general 5-parameter subfamily of the MMR
spacetime representing a pair of KN black holes with a separating
strut, and our objective is to add a nonzero net charge $2q$ to
the solution considered in the previous section and get the
general expression for $\s$ in (\ref{sig1}) in terms of $q$ and
$\beta$. Note that the case of two KN black holes with equal
electric and opposite magnetic arbitrary charges has not been
considered before and it represents a physically and
mathematically nontrivial example of a binary dyonic
configuration.

To fulfil our goal, we must first reparametrize the entire MMR
solution using the transformation formulas (\ref{sub1}) and
(\ref{sub2}). The expressions of the Ernst potentials $\E$ and
$\Phi$ thus obtained are the following:
\bea \E&=&(A-B)/(A+B), \quad \Phi=C/(A+B), \nonumber\\
A&=&R^2\{[(R^2+4a^2)(\s^2-a^2)-4(m^2-q^2)(2\s^2-m^2+q^2)
+4a(qb+m\delta)] \nonumber\\ &&\times(R_+r_-+R_-r_+)
-2i\s[a(R^2-4\Delta)-2(qb+m\delta)](R_+r_--R_-r_+)\} \nonumber\\
&&+4\s^2\{[2(m^2-q^2)(R^2-2m^2+2q^2)-(R^2-4a^2)
(\s^2+a^2)-4a(qb+m\delta)] \nonumber\\ &&\times(R_+r_++R_-r_-)
+2iR[2a(\s^2-\Delta)-qb-m\delta](R_+r_+-R_-r_-)\} \nonumber\\
&&+(R^2-4\s^2)[(R^2+4a^2)(\s^2+a^2)-4(m^2-q^2)^2
-4a(qb+m\delta)](R_+R_-+r_+r_-), \nonumber\\
B&=&2R\s\{(R^2-4\s^2)[mR\s(R_++R_-+r_++r_-) -(2m\Delta+a\delta)
(R_++R_--r_+-r_-)] \nonumber\\ &&
+i[ma(R^2+4\s^2-8\Delta)-4mqb-2\delta(2m^2-\Delta)][(R-2\s)(R_--R_+) \nonumber\\
&&+(R+2\s)(r_+-r_-)]  +iR\s\delta
[(R-2\s)(R_--R_+)-(R+2\s)(r_+-r_-)]\}, \nonumber\\
C&=&(q/m)B+2R(\s/m)(mb+q\delta)[a(R^2-4\s^2)(R_++R_--r_+-r_-)
-2iR(\s^2-\Delta) \nonumber\\
&&\times(R_+-R_--r_++r_-) +i\s(R^2-4\Delta)(R_+-R_-+r_+-r_-)], \nonumber\\
R_\pm&=&\sqrt{\rho^2+\left[z\pm({\textstyle{\frac{1}{2}}}R+\s)\right]^2},
\quad
r_\pm=\sqrt{\rho^2+\left[z\pm({\textstyle{\frac{1}{2}}}R-\s)\right]^2},
\quad \Delta\equiv m^2-a^2-q^2, \label{EF3} \eea
and formulas (\ref{MF1}) for the metric functions take the form
\bea f&=&\frac{A\bar A-B\bar B+C\bar C}{(A+B)(\bar A+\bar B)},
\,\,\, e^{2\gamma}=\frac{A\bar A-B\bar B+C\bar C}{16R^4\s^4
(R^2-4\s^2)^2R_+R_-r_+r_-}, \,\,\, \omega=-\frac{{\rm Im}
[G(\bar A+\bar B)+C\bar I]}{A\bar A-B\bar B+C\bar C}, \nonumber\\
G&=&4R\s^2[(R^2-4\s^2)(2m^2-q^2) +b^2-2\delta^2]
(R_-r_--R_+r_+)+2R^2\s\{
[(R^2-4\s^2) \nonumber\\
&& \times(2m^2-q^2) -b^2+2\delta^2] (R_-r_+-R_+r_-)
+8i\s(bq+2m\delta)(R_+-R_-)(r_+-r_-)\} \nonumber\\
&& -2i[a(2m^2-q^2)(R^2+4\s^2-8\Delta)-4a^2(qb+2m\delta)
-4mq(mb+q\delta)-a(b^2-2\delta^2)] \nonumber\\
&&\times[R^2(R_+-r_+)(R_--r_-)-4\s^2(R_+-r_-)(R_--r_+)]-2R(\s/m)
\nonumber\\ &&\times \{[m^2(R^2+4\s^2-8\Delta)-\delta(4ma-\delta)]
[\s(R^2-4\Delta)(R_--R_++r_--r_+) \nonumber\\
&&-2R(\s^2-\Delta)
(R_--R_+-r_-+r_+)+ia(R^2-4\s^2)(R_++R_--r_+-r_-)] \nonumber\\
&& -2mb(mb+q\delta)[(R-2\s)(R_--R_+)-(R+2\s)(r_--r_+)]
-2imq(R^2-4\s^2) \nonumber\\ &&\times(mb+q\delta)(R_-+R_+-r_--r_+)
\}
-2zB+i(4ma-\delta)B/m-2qC, \nonumber\\
I&=&2R\s\{R[mq(R^2-4\s^2)-b\delta](R_-r_+-R_+r_-)
-2\s[mq(R^2-4\s^2)+b\delta](R_+r_+-R_-r_-) \nonumber\\
&&+i(R^2-4\s^2)(mb+q\delta)(R_+R_- -r_+r_-+4R\s)\}
-(i/2)(R^2-4\s^2)(mb-q\delta) \nonumber\\
&&\times[R^2(R_+r_-+R_-r_+) +4\s^2(R_+r_++R_-r_-)]
+(i/2)[(R^2+4\s^2-8\Delta)-4ab\delta \nonumber\\
&&-16mq(qb+m\delta)] [R^2(R_+r_-+R_-r_+) -4\s^2(R_+r_++R_-r_-)]
-2i(R^2-4\s^2) \nonumber\\
&&\times[maq(R^2+4\s^2-8\Delta)-2q\delta(2m^2-\Delta)
-2mb(m^2-a^2+q^2)-ab\delta] (R_+R_-+r_+r_-)\nonumber\\
&&-R(\s/m)\{2R[(mq(R^2+4\s^2-8\Delta)+b(4ma-\delta))(\Delta-\s^2)
+4m\delta(mb+q\delta)] \nonumber\\ &&\times(R_--R_+-r_-+r_+)
+\s[(mq(R^2+4\s^2-8\Delta)+b(4ma-\delta))(R^2-4\Delta) \nonumber\\
&&-16m\delta(mb+q\delta)](R_--R_++r_--r_+) -i(R^2-4\s^2)
[8m^2(mb+q\delta)-ab(4ma-\delta) \nonumber\\
&&-maq(R^2+4\s^2-8\Delta)](R_-+R_+-r_--r_+)\}+2q(A+B)
+ibB/(2m)-(z-2ia)C. \nonumber\\ \label{MF3} \eea
As before, the electromagnetic potentials $A_t$ and $A_\varphi$
are determined by formulas (\ref{At}).

In the presence of the strut, which means that $\s$ is not
arbitrary but verifies (\ref{sig1}), the parameter $q$ is the
electric charge of each KN black hole, while the magnetic charge
$\beta$ must be introduced by means of the relation of the
magnetic dipole parameter $b$ to the charges $q$ and $\beta$. Such
a relation turns out to be slightly more complicated than in the
pure magnetic case considered in the previous section, and it can
be written as
\be b=-\frac{(R^2-4\Delta)(2aq+R\beta+4q\mu)}
{R^2+2mR+4a^2+8a\mu}, \quad \mu\equiv\frac{a(q^2-\beta^2)
+q\beta(R+2m)}{(R+2m)^2+4a^2}. \label{b2} \ee
Then after the substitution of (\ref{b2}) into (\ref{sig1}) and
choosing the minus sign we get the desired final formula for $\s$,
namely,
\be \s=\sqrt{m^2-a^2-q^2+\frac{(R^2-4\Delta)
\{4[ma+(R+2m)\mu]^2-(2aq+R\beta+4q\mu)^2\}}
{(R^2+2mR+4a^2+8a\mu)^2}}, \label{sig3} \ee
while the expression for $\delta$ obtainable from (\ref{delta1}),
(\ref{b2}) and (\ref{sig3}) has the form
\be \delta=\frac{2(R^2-4\Delta)[ma+(R+2m)\mu]}
{R^2+2mR+4a^2+8a\mu}. \label{delta3} \ee

The angular momentum of each black hole is now defined by the
expression
\be J=\frac{[(R+2m)^2+4a^2][ma-(R-2m)\mu] -4q^2[ma+(R+2m)\mu]}
{R^2+2mR+4a^2+8a\mu}, \label{AM2} \ee
so that the two black holes have the same mass $m$, angular
momentum $J$ and electric charge $q$, but they differ in their
magnetic charges: $\beta$ of the lower and $-\beta$ of the upper
black hole (see Fig.~2). Therefore, we have a nontrivial binary
system of dyonic KN black holes in which the magnetic charges are
not introduced via the duality rotation of the potential $\Phi$,
in contrast to all the dyonic solutions studied for example in the
paper \cite{MGa}. In what follows we shall see that thermodynamics
of the black holes in our system is subject to the generalized
Smarr mass formula which takes into account the contribution of
the magnetic field.

The calculations performed for the lower black hole with the aid
of the standard Tomimatsu's formulas \cite{Tom} give for the
entropy, Hawking temperature, horizon's angular velocity and the
electric potential the following expressions:
\bea S&=&\frac{\s}{2T}=
\frac{\pi\{[mR+(R+2m)\s+2\Delta]^2+[a(R+2\s)+\delta]^2\}}
{R(R+2\s)}, \nonumber\\
\Omega&=&\frac{a[R^2-4\Delta+2\s(R+2\s)]-2(qb+m\delta)}
{[mR+(R+2m)\s+2\Delta]^2+[a(R+2\s)+\delta]^2}, \nonumber\\
\Phi_e&=&\frac{q(R+2\s)[mR+(R+2m)\s+2\Delta]-b[a(R+2\s)+\delta]}
{[mR+(R+2m)\s+2\Delta]^2+[a(R+2\s)+\delta]^2}, \label{TQ2} \eea
and we have used the same way of writing these quantities as in
the paper \cite{CCH}.\footnote{Note that formulas (\ref{TQ2}) are
valid in the case of the general MMR solution, independently of
the existence of a strut, and hence in principle need further
processing to introduce explicitly the magnetic charge parameter
$\beta$. However, this way of writing the thermodynamical
quantities permits one to see a little bit better the mathematical
structure of the potential $\Phi_m$ which we introduce later on
and its relation to other thermodynamic characteristics.}

The above formulas must be supplemented with the expression of the
magnetic potential $\Phi_m$ which, according to the papers
\cite{Tom,MGa}, is defined by the equation
\be \beta\Phi_m=-\frac{1}{2}\int_{H}(A_\varphi A'_\varphi)_{,z} d
z, \label{eq_FM} \ee
where $A'_\varphi={\rm Im}(\Phi)$. Remarkably, the magnetic
potential $\Phi_m$ can be written in a concise form
\be \Phi_m=-\frac{[2q\delta+b(R+2m)]\{
q[a(R+2\s)+\delta]+\beta[mR+(R+2m)\s+2\Delta]\}}
{[2aq+\beta(R+2m)]\{
[mR+(R+2m)\s+2\Delta]^2+[a(R+2\s)+\delta]^2\}}, \label{FHM} \ee
and it is not difficult to check that in the absence of electric
charge ($q=0$) formula (\ref{FHM}) reduces to the expression of
$\Phi_m$ in (\ref{TQ1}).

The thermodynamical variables obtained verify the generalized mass
relation
\be m=2TS+2\Omega J+\Phi_e q+\Phi_m\beta, \label{Sma_g} \ee
and it should be remarked that the same relation holds for the
upper black hole because the integral on the right-hand side of
(\ref{eq_FM}) gives the same result as for the lower black hole,
which must be interpreted as changing the sign of $\Phi_m$ (and
$\beta$) on the upper horizon, while all other thermodynamical
quantities remain unchanged.

To write out the corresponding first law of thermodynamics for our
binary system, we still need the expressions of the interaction
force and thermodynamic length. The calculations give for the
former quantity the expression
\bea {\cal F}&=&\frac{(R^2+4mR+4\Delta)(m^2-q^2-4\mu^2)
+4q[q\Delta+R\beta(a+2\mu)]+R^2\beta^2+4a\mu(R^2-4\Delta)}
{(R^2-4\Delta)[(R+2m)^2+4a^2]}, \nonumber\\ \label{F2} \eea
while the latter quantity was found to have the form
\be \ell=\frac{(R^2-4\Delta)^2[(R+2m)^2+4a^2]}
{(R+2\s)(R^2+2mR+4a^2+8a\mu)^2}. \label{el2} \ee
Then the first law reads as follows:
\be dM_T=2TdS+2\Omega dJ+2\Phi_e dq+2\Phi_m d\beta-\ell d{\cal F},
\quad M_T=2m, \label{FL1} \ee
and here both the electric and magnetic contributions are taken
into account consistently. At the same time, while the potentials
$\Phi_e$ and $\Phi_m$ are symmetric with respect to the change
$q\to\beta$, $\beta\to q$ in the solutions where the magnetic
charges are introduced by means of the duality rotation of the
Ernst potential $\Phi$ \cite{MGa}, in our nontrivial dyonic
configuration these $\Phi_e$ and $\Phi_m$ are defined by
different, nonsymmetric expressions. As a consequence, the
generalized Smarr formula (\ref{Sma_g}) in our case cannot be cast
into a more elegant form (by introducing a complex charge
$q+i\beta$) like this was done in the paper \cite{MGa}.

\section{Discussion}

Since the MMR solution is the general equatorially symmetric
2-soliton solution of the stationary axisymmetric electrovac
problem (of course up to an arbitrary duality rotation of the
electromagnetic potential $\Phi$ \cite{EMR}) then its 5-parameter
subfamily considered in the previous section can be viewed as
describing the general configuration of two identical corotating
black holes with a massless strut in between. This in turn means
that any known exact solution for a binary system with equatorial
symmetry must be a particular specialization of the latter
subfamily or obtainable from it via the constant phase
transformation $\exp(i\a)$ of the potential $\Phi$. In this
respect, the recent solutions for corotating KN black holes with
identical or opposite electric charges considered in \cite{CCH}
belong to our 5-parameter subfamily because the first solution is
just its $\beta=0$ particular case, while the second one follows
immediately from its $q=0$ specialization by applying the constant
phase transformation with $\a=\pi/2$. The dyonic generalizations
of the solutions \cite{CCH} performed in \cite{Cab} are also
trivially obtainable from the $\beta=0$ and $q=0$ specializations
of our subfamily. Note that the main physical difference between
our nontrivial dyonic solution and those presented in \cite{Cab}
is that the latter solutions become static in the absence of the
rotation parameter $a$, while the former solution at $a=0$ still
remains stationary due to the well-known frame-dragging effect by
a charged magnetic dipole \cite{Das,Bon}.

The 5-parameter dyonic configuration defined by formulas
(\ref{b2})-(\ref{AM2}) has proved to be a good example of a binary
system whose thermodynamics is subject to the generalized Smarr
formula which takes into account the contribution of magnetic
charges. It may be recalled in this regard that the recent paper
\cite{CGa} has argued that the magnetic potential $\Phi_m$ should
not arise in the Smarr mass relation, which would mean in
particular that the latter relation, say, for the magnetically
charged KN black hole must look like in the case of an uncharged
Kerr black hole. Though the constructions of the paper \cite{CGa}
were already shown to be frankly unphysical and inconsistent
\cite{GMR}, a recent preprint \cite{Cab} still makes an effort to
rehabilitate the results of the paper \cite{CGa} through the
analysis of a specific binary dyonic configuration of KN black
holes. The main contradiction of the author of \cite{Cab} is that
he starts with the mass relation without the magnetic potential
$\Phi_m$ (like in the paper \cite{CGa}) but eventually, after some
manipulations, arrives at the Smarr formula of the paper
\cite{MGa} in which the potential $\Phi_m$ is already present,
thus fully ignoring that precisely his final result was the
subject of criticism in the paper \cite{CGa}. We hope that our
analysis of the first law of thermodynamics carried out in the
previous section confirms convincingly the correctness of the
original Tomimatsu's vision of the Smarr mass formula.

An intriguing aspect of the binary charged black hole
configurations which is of special interest to us and which we
would like to briefly comment here is the following. In our papers
on the binary systems of identical Kerr sources \cite{MRu3,MRu4}
we have shown that the uniqueness of the binary configurations
with fixed masses and angular momenta can be broken for some
particular values of the parameters, so that up to three different
configurations with the same masses and angular momenta may exist
due to relation of the rotation parameter $a$ to the individual
angular momentum $J$ via the cubic equation. Since the analogous
relation of the parameter $a$ to $J$ in the formulas (\ref{AM1})
and (\ref{AM2}) is determined, as can be easily seen, by a quintic
equation, a natural question arises of whether the electromagnetic
field of KN black holes is able to increase the nonuniqueness in
the binary systems of charged black holes up to five different
configurations with the same masses, angular momenta and charges?
Our first numerical examination of equations (\ref{AM1}) and
(\ref{AM2}) has not yet been able to detect the parameter sets at
which these equations would get five real roots for $a$. In the
majority of cases these equations have one real root and two pairs
of complex conjugate roots, they also may have three real and two
complex roots. In the latter case a situation is possible when in
the initial parameter sets ensuring three real roots of equations
(\ref{AM1}) and (\ref{AM2}) the subsequent increase of the values
of $q$ and $\beta$ (keeping $m$ and $J$ unchanged) leads to
disappearance of two real roots, thus getting unique
configurations from nonunique ones. Anyway, should any particular
parameter sets at which the above quintic equations admit five
real roots exist, they must belong to a highly restricted sector
of the parameter space which yet has to be identified in the
future.

\section*{Acknowledgments}

We are grateful to Hugo Garc\'ia-Compe\'an and Carlos
Ram\'irez-Valdez for many interesting and helpful discussions on
thermodynamics of charged black holes. This work was partially
supported by CONACyT of Mexico, by Project PGC2018-096038-B-100
from Ministerio de Ciencia, Innovaci\'on y Universidades of Spain,
and by Project SA083P17 from Junta de Castilla y Le\'on of Spain.

\newpage

\begin{figure}[htb]
\centerline{\epsfysize=75mm\epsffile{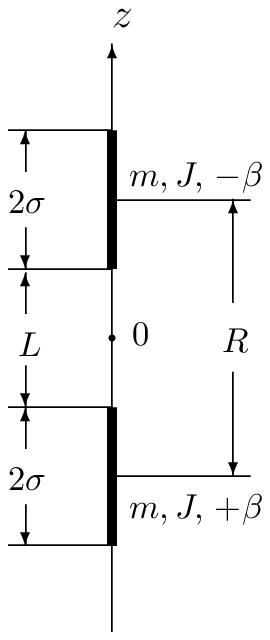}} \caption{Location
of two equal corotating KN black holes with opposite magnetic
charges on the symmetry axis. $L=R-2\s$ is the coordinate length
of the strut.}
\end{figure}

\begin{figure}[htb]
\centerline{\epsfysize=75mm\epsffile{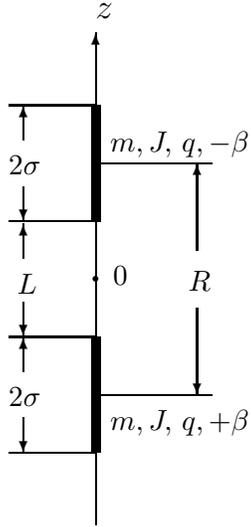}} \caption{Location
of two equal corotating KN black holes with equal electric and
opposite magnetic charges on the symmetry axis. $L=R-2\s$ is the
coordinate length of the strut.}
\end{figure}

\end{document}